\documentclass[conference]{IEEEtran}
\IEEEoverridecommandlockouts
\usepackage{cite}
\usepackage{amsmath,amssymb,amsfonts}
\usepackage{algorithmic}
\usepackage{graphicx}
\usepackage{textcomp}
\usepackage{xcolor}
\usepackage[hidelinks]{hyperref}
\usepackage{caption}
\usepackage{subcaption}
\usepackage{gensymb}
\usepackage{float}
\usepackage{array}
\usepackage{physics}
\usepackage{optidef}
\usepackage{algorithm}
\usepackage{algorithmic}
\usepackage{bm}

\usepackage{url}
\def\BibTeX{{\rm B\kern-.05em{\sc i\kern-.025em b}\kern-.08em
    T\kern-.1667em\lower.7ex\hbox{E}\kern-.125emX}}
\begin{document}

\title{Analyzing Cross-Phase Effects of Reactive Power Intervention on Distribution Voltage Control}

\author{\IEEEauthorblockN{Dhaval Dalal, \textit{Senior Member, IEEE}, 
 Anamitra Pal, \textit{Senior Member, IEEE}, and  Raja Ayyanar, \textit{Fellow, IEEE}}
\IEEEauthorblockA{\textit{School of Electrical, Computer and Energy Engineering} \\
\textit{Arizona State University, Tempe, AZ, USA}\\
ddalal2@asu.edu, anamitra.pal@asu.edu, rayyanar@asu.edu}
\thanks{This work was supported in part by the US Department of Energy under grant DE-EE0009355.}}
\maketitle

\begin{abstract}
Increasing photovoltaic (PV) penetration in the distribution system can often lead to voltage violations. Mitigation of these violations requires reactive power intervention from PV inverters. However, the unbalanced nature of the distribution system leads to mixed effects on the voltages of nearby nodes for each inverter injecting or absorbing reactive power. In particular, reactive power absorption to reduce over-voltage in one phase can exacerbate over-voltage in a different phase.
In this paper, the factors impacting the incremental and decremental voltage effects of reactive power intervention are analyzed in detail. The result of these effects on the distribution system performance is presented to highlight their significance and the need to factor them in for any coordinated voltage control algorithm. 
\end{abstract}


\section{Introduction}
The road to net zero emissions (NZE) has milestones for achieving significantly high level of photovoltaic (PV) penetration in the distribution systems \cite{3x}. However, due to limited planning control and operational visibility from the utility
side, the proliferation of PVs can lead to potential voltage violations which negatively impact the distribution system quality and reliability \cite{4x}. The contributors to voltage violations include a combination of factors such as power flow reversal, low X/R ratio, inappropriate settings of step voltage regulators (SVRs) and capacitor banks (CBs), and sudden drastic changes in PV generation and load profiles \cite{9x}. 

Two major avenues for addressing the voltage violations are: (a) on-load-tap-changer (OLTC)/SVR/CB control, or (b) reactive power (Q) intervention (e.g., injection or absorption) from PV inverters. A combination of the two provides a third option. OLTC/SVR/CB control has severe maintenance cost penalty if implemented to address the diurnal variations in PV generation. 
Hence, the focus of current research is
on inverter Q control. Localized voltage control techniques such as inverter volt-VAr (VV) control require no coordination, but have limited effectiveness when the PV penetration is high
\cite{7x}. Communication-based methodologies, in which the inverters are actively monitored and commanded in a coordinated fashion using information not locally available to each inverter, are more effective for PV-rich systems \cite{Lusis, Yao}. 
Since coordinated control on large distribution systems have significant communication and computation overhead, a number of recent works have proposed decentralized voltage regulation by partitioning the distribution networks into zones \cite{Zon_1,Zon_2}. 
However, any partitioning of the system results in certain parametric effects being ignored, because they cannot transcend the partition boundaries. Specifically, the unbalanced nature of distribution systems create significant cross-phase voltage effects, often leading to new voltage violations when Q intervention is used in a single-phase PV inverter. Since many of the zoning approaches rely on phase-based partitions (e.g., \cite{Zon_1,Zon_2,Zon_3}), the cross-phase effects are disregarded, resulting in sub-optimal outcomes.

In this paper, a methodical approach is developed to characterize the  cross-phase voltage effects caused by Q intervention, including their causes, sensitivities, and end-effects on system performance. 
The value of considering these effects is demonstrated by testing on a renewable-rich complex feeder\footnote{A complex feeder is one that has both primary and secondary circuits, unbalanced multi-phase lines/loads, and OLTCs/SVRs/capacitors.} with and without these effects factored in. 

The major contributions of this paper are as follows: 
\begin{itemize}
    \item Identification of cross-phase voltage effects as a major impediment in achieving desired voltage regulation in distribution networks with high PV penetration
    \item Sensitivity analyses to identify and quantify the factors impacting cross-phase voltage effects
    \item Proposing and validating
    a novel methodology to integrate cross-phase voltage effects into a voltage regulation algorithm to achieve higher hosting capacity (HC)
\end{itemize}

\section{Characteristics of Unbalanced Distribution Systems}
Distribution systems are significantly different from transmission systems in terms of power flow balance between phases, line impedances, and/or types of distributed energy resources (DERs). 
The role of these factors in maintaining reliable system operation 
is well understood 
for traditional load-dominant distribution systems or systems with low levels of DER penetration \cite{kersting}. 
However, there is a need to investigate the behavior of distribution systems with a high percentage of DERs that are participating in voltage regulation by injecting or absorbing significant levels of Q. Following characteristics of distribution systems merit special attention:
\begin{enumerate}
    \item \textbf{Low X/R ratios:} Smaller conductor sizes in the distribution system lead to much larger increases in the per unit resistance values ($ \propto \frac{1}{d^2}$) compared to the increases in (self and mutual) inductances ($\propto \log \frac{1}{d}$). For instance, X/R ratios in many distribution systems are in between 2 and 4 as compared to $>$10 for transmission systems. 
    \item \textbf{Diversity of line configurations:} Distribution systems consist of single-phase, two-phase, and untransposed three-phase lines serving unbalanced loads. Therefore, a positive sequence analysis approach is not valid. As stated in \cite{kersting}, ``it is necessary to retain the identity of the self- and mutual impedance terms of the conductors in addition to taking into account the ground return path for the unbalanced currents".
    \item \textbf{DER types:} 
    The addition of behind-the-meter DERs (typically, PVs) is not controlled by the utilities. They are likely to add to the phase imbalance caused by the loads, while also adding high degrees of uncertainty and  variability. Particularly, the PV inverters have varying capability to inject or absorb Q. 
\end{enumerate}

In summary, when voltage mitigation is done in PV-rich systems by providing Q absorption/injection, the imbalance and low X/R ratios play a major role in determining the outcome of the Q intervention \cite{Niknam}.

\section{Simple Distribution System Illustration} 
\label{Simple_Ch}

\begin{figure}[htbp]
\centering
   \includegraphics[height =3.5 cm, keepaspectratio]{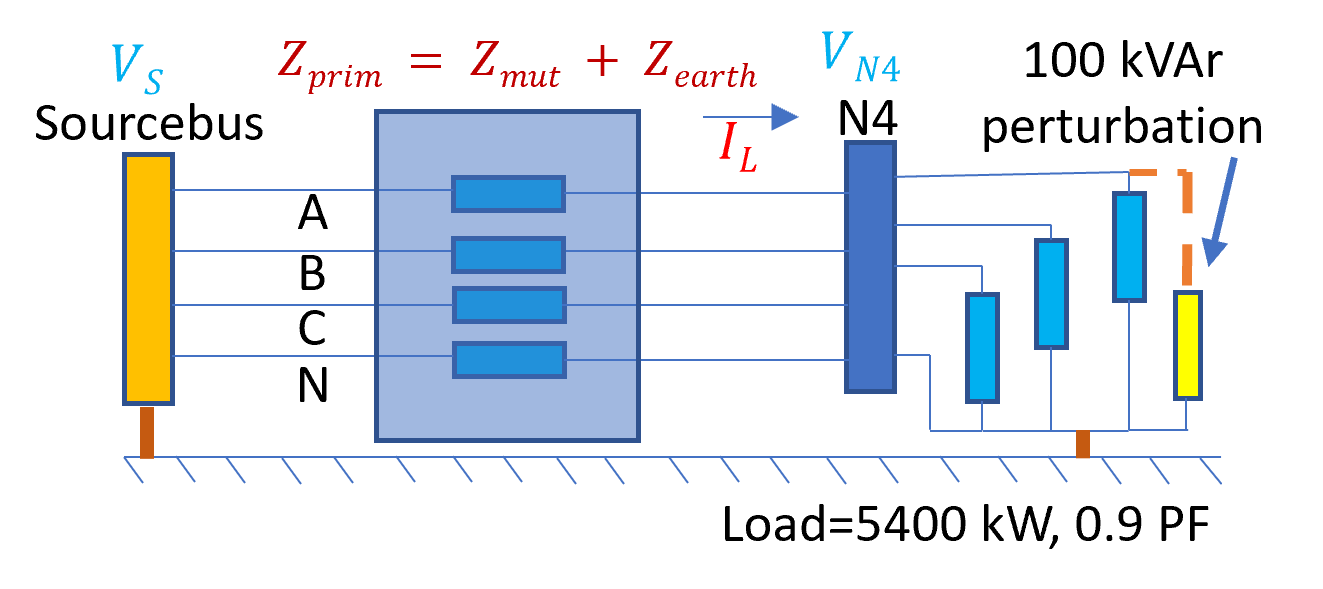}
  \vspace{-1em}
  \caption{Simple 4-wire, 2-bus system}
  \label{fig: 4wire}
\end{figure}

To fully understand the effects of Q intervention in distribution networks, it is useful to first construct a simple 4-wire, 2-bus system as shown in Fig. \ref{fig: 4wire}. 
Note that this system is a reduced version of the 4-bus test case available in OpenDSS, with the transformer removed and the feeder-head line-to-line voltage changed to 4.16 kV. All other configuration details including wire data, line geometry, and load were retained from the original case. The  system has a balanced 3-phase load (5400 kW, 0.9 power factor (PF)), representing the base case ($i=0$).  A primitive impedance matrix, $Z_{\mathrm{prim}} \in \mathbb{R}^{4 \times 4}$,
consisting of $Z_{\mathrm{mut}}$ and $Z_{\mathrm{earth}}$ components was constructed from the available data using modified Carson's equations. With the given parameters, the X/R ratio is about 3:1. The power flow equations are given by:
\begin{equation}\label{S_eq}
    \mathbf{S_i} = \mathbf{P_i} + j \mathbf{Q_i}
\end{equation}
\begin{equation}\label{IL_eq}
    \mathbf{I_{L_i}} = \Bigg(\frac{\mathbf{{S_i}}}{\mathbf{V_{{N4}_i}}}\Bigg)^*
\end{equation}
\begin{equation}\label{V_eq}
    \mathbf{V_{{N4}_i}} = \mathbf{V_S} - \mathbf{I_{L_i}}  \times Z_{\mathrm{prim}}
\end{equation}
where, $\mathbf{S_i}$, $\mathbf{P_i}$, and $\mathbf{Q_i}$ are the load apparent, real and reactive power  vectors, $\mathbf{I_{L_i}}$ is the complex vector representing the line current, $^*$ in the superscript denotes the conjugate operation, $\mathbf{V_S}$ is the base voltage vector (a constant), and $\mathbf{V_{{N4}_i}}$ is the voltage vector at the load node. 
It was observed that despite a balanced load, the $\mathbf{V_{{N4}_0}}$ values exhibited imbalance, which is attributable to the mutual impedance mismatches related to wire location differences. 
The results obtained by solving \eqref{S_eq}-\eqref{V_eq} numerically
were found to be consistent with those obtained from the OpenDSS solution of this system.
The phase voltage magnitudes are shown in the first row of Table \ref{tab: 4W_res}.

Next, a perturbation of 100 kVAr ($Q_{1_a} = Q_{0_a} + 100$ kVAR) is applied to phase A to test the effects of PV inverter Q absorption ($i=1$). Solutions to \eqref{S_eq}-\eqref{V_eq} were again found to be consistent with the OpenDSS results. The phase voltage magnitudes are provided in the second row of Table \ref{tab: 4W_res}. The voltage difference between the two cases is shown in the third row of the table. Since the objective of Q absorption is to reduce the voltage, the results for phase A are consistent with that objective. However, it is observed that both phase B and phase C have voltage differences of opposite polarity. 

\begin{table}[htbp]
\centering
\caption{N4 phase voltages (in Volts)}
\begin{tabular}{|c|c|c|c|c|}
\hline
\textbf{Case} & \textbf{Phase A}& \textbf{Phase B}& \textbf{Phase C}  \\
\hline
Base ($i=0$) & 2103 & 2206 & 2150  \\
\hline
+100 kVAr ($i=1$) & 2068 & 2217 & 2160  \\
\hline
Difference  & -35.3 & 11.6 & 10.0  \\
\hline
Difference due to $Z_{\mathrm{earth}}$  & 19.2 $\angle180 \degree$ & 19.2 $\angle180 \degree$ & 19.2 $\angle180 \degree $ \\
\hline
Difference due to $Z_{\mathrm{mut}}$  & 16.4 $\angle165 \degree$ & 2.1 $\angle -57 \degree$ & 3.4 $\angle 57 \degree$ \\
\hline
\end{tabular}
\label{tab: 4W_res}
\end{table}

Before providing contextual analysis of these results, it is instructive to
find the 
contributions 
to the voltage differences
from $Z_{\mathrm{earth}}$ ($\mathbf{\Delta V_{{N4}_e}}$) and $Z_{\mathrm{mut}}$ ($\mathbf{\Delta V_{{N4}_m}}$), respectively.
Note that this breakdown is not available from the OpenDSS power flow results,
but can be obtained by solving the following equations that
decompose the contributions. 
The dominant component of $\mathbf{\Delta I_{L}}$ is the imaginary component of phase A. It leads to the computed values of $\mathbf{\Delta V_{{N4}_e}}$ and $\mathbf{\Delta V_{{N4}_m}}$ from \eqref{V_earth} and \eqref{V_mult}, respectively, which are shown in the last two rows of Table \ref{tab: 4W_res}.
%

\begin{equation}\label{IL_diff}
    \mathbf{\Delta I_{L}} = \mathbf{I_{L_1}} - \mathbf{I_{L_0}}
\end{equation}
\begin{equation}\label{V_earth}
    \mathbf{\Delta V_{{N4}_e}} = \mathbf{\Delta I_{L}} \times Z_{\mathrm{earth}}
\end{equation}
\begin{equation}\label{V_mult}
    \mathbf{\Delta V_{{N4}_m}} = \mathbf{\Delta I_{L}} \times Z_{\mathrm{mut}}
\end{equation}

\begin{figure}[b!]
\centering
\begin{subfigure}[t] {0.485\textwidth} 
    \centering
    \includegraphics[width=1\linewidth]{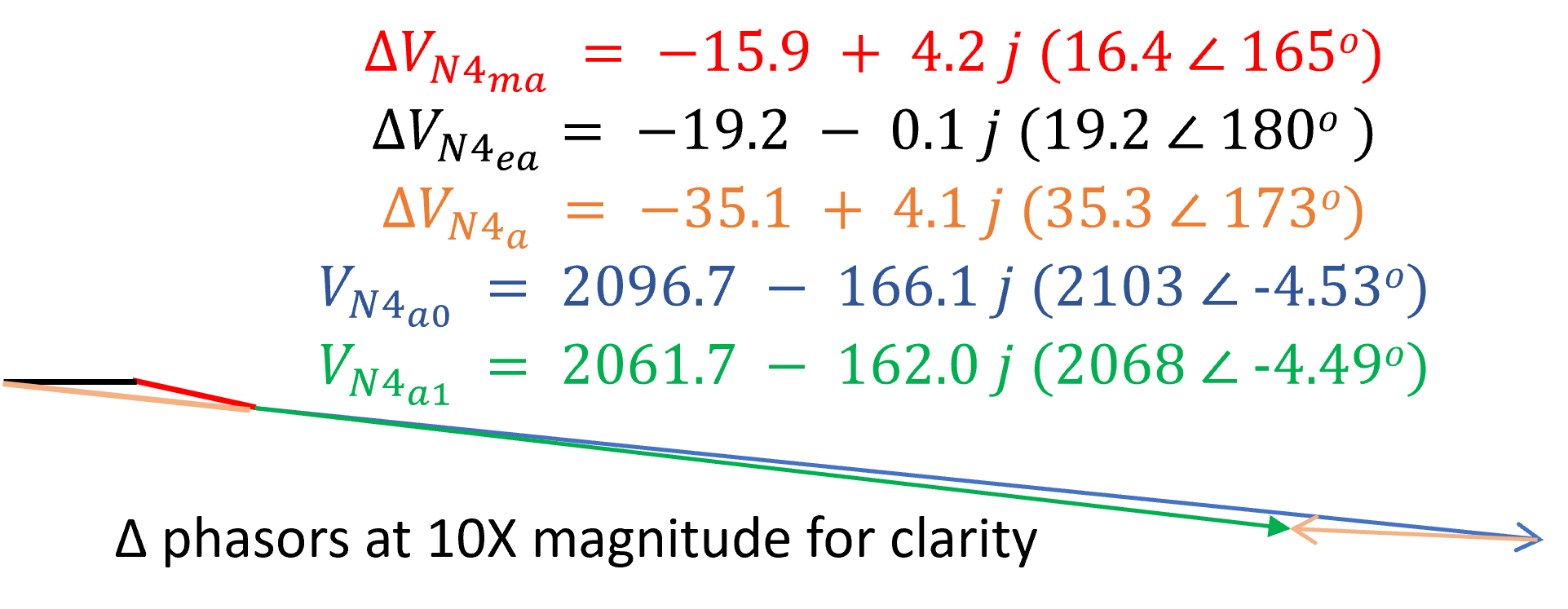}
    \caption{Phase A}
  \label{fig: PhasorA }
\end{subfigure}
\hfill
\begin{subfigure}[h] {0.485\textwidth}
    \centering
    \includegraphics[height =5.5 cm, keepaspectratio]{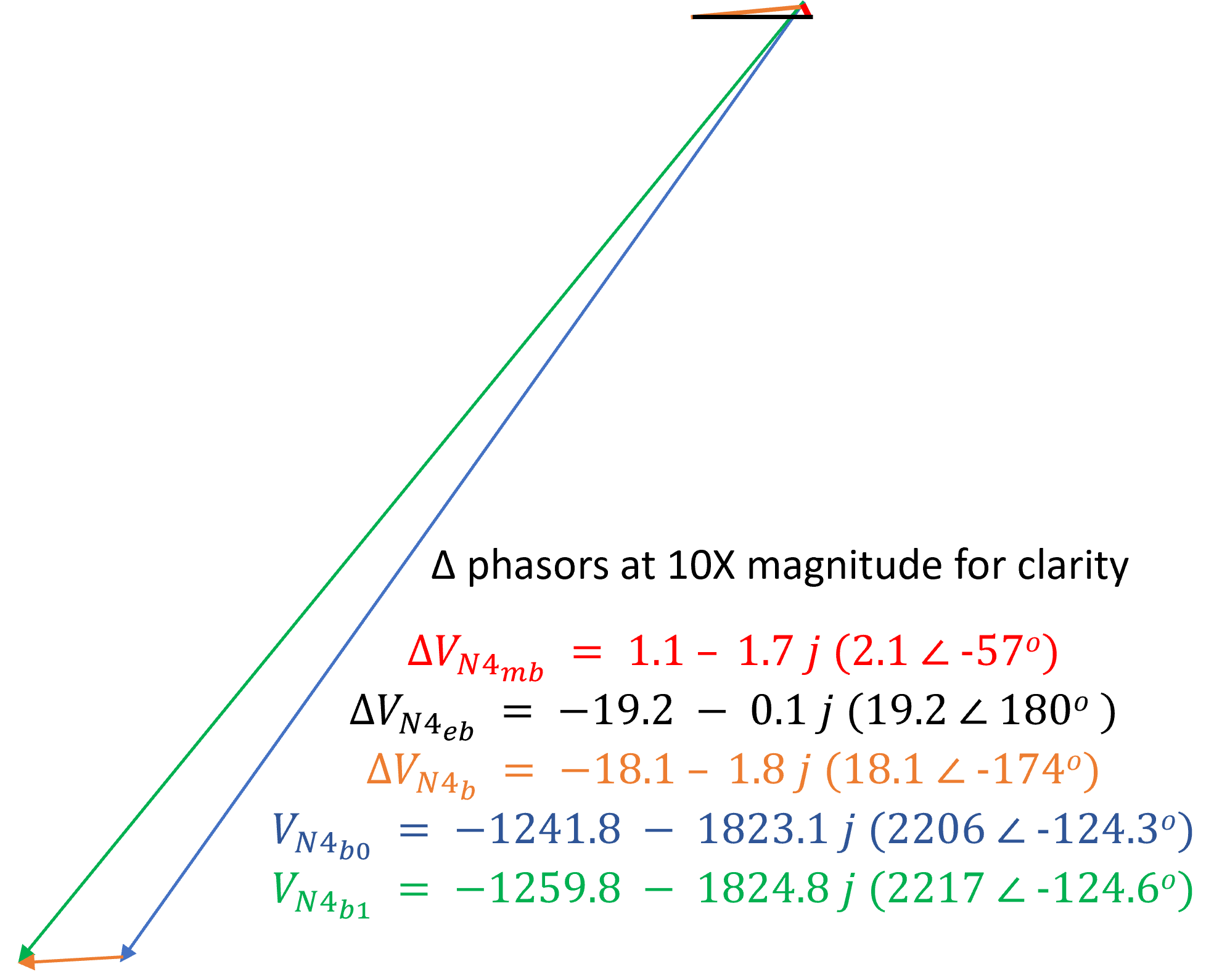}
    \caption{Phase B}
  \label{fig: PhasorB}
\end{subfigure}    
\begin{subfigure}[h] {0.485\textwidth}
    \centering
    \includegraphics[height =5 cm, keepaspectratio]{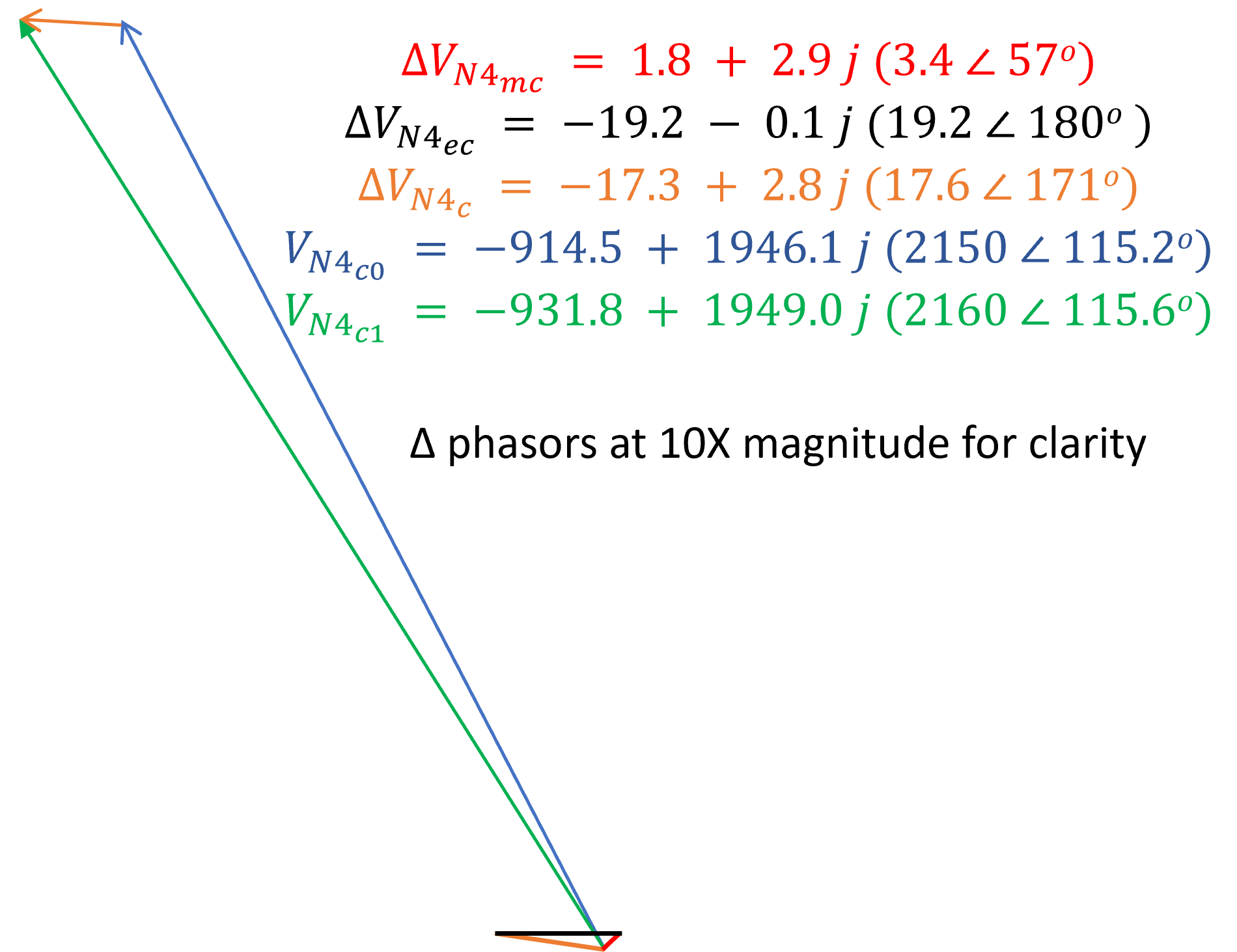}
    \caption{Phase C}
  \label{fig: PhasorC}
\end{subfigure}   
    \caption{Phasor diagrams showing effects of Q perturbation}
  \label{fig: Phasor}
\end{figure}

The results in Table \ref{tab: 4W_res} provide significant insights into the incremental and decremental effects of changes in reactive power ($\Delta Q$). The positive voltage deviations in phases B and C as a result of Q absorption in phase A show unintended consequences of a corrective action (i.e., voltage mitigation for phase A) that must be recognized. Additionally, the decomposition of the contributions from $Z_{\mathrm{mut}}$ and $Z_{\mathrm{earth}}$ show that the corrective effect for phase A has nearly equal and significant contributions from both factors, while the phase B and phase C effects are dominated by the $Z_{\mathrm{earth}}$ factor (compare the entries in the last two rows of Table \ref{tab: 4W_res} for phases B and C). 
To understand why the earth current has such a large opposite-polarity effect on the other phases, we look at the phasor diagrams of the voltages in Fig. \ref{fig: Phasor}. Individual phasors are color coded and their values are shown next to the diagrams for clarity.  

For phase A, the contributions from $Z_{\mathrm{mut}}$ and $Z_{\mathrm{earth}}$ have similar phase and magnitudes, and the combination of the two results in phase A voltage reduction (which is expected). 
Since the earth current is the same, the $\mathbf{\Delta V_{{N4}_e}}$ is identical for all three phases, as depicted by the vectors in black color
and values in the phasor diagrams. Moreover, since the perturbation was made in phase A, the effects of $Z_{\mathrm{mut}}$ on phases B and C voltages are minimal as shown by the red-colored vectors
and values. As a result, the composite effect
of $\mathbf{\Delta V_{N4}}$ for phases B and C is dominated by the $\mathbf{\Delta V_{{N4}_e}}$ component, which is 19.2 V at -180\degree. As seen in Figs. \ref{fig: PhasorB} and \ref{fig: PhasorC}, the phasor addition of $\mathbf{\Delta V_{N4}}$ and $\mathbf{V_{{N4}_0}}$ for phases B and C result in higher magnitudes of $\mathbf{V_{{N4}_1}}$ with a very small change in angle. 

While these results allow a good analytical basis for understanding the cross-phase effects and their causes, they need to be put into proper context. Firstly, the reported voltage deviations are fairly small in p.u. terms (-0.015 p.u. for phase A, and about 0.005 p.u. for phases B and C). Secondly, a real distribution system rarely has the 4-wire line configuration shown in Fig. \ref{fig: 4wire} with a single-phase PV on one of the lines. Hence, the question to address is whether actual distribution systems need to contend with both incremental and decremental effects when using Q intervention from DERs as a voltage control technique. 
This is analyzed in the next section.

\section{Evaluation on a Complex Distribution System}
For this purpose, we evaluate the J1 Feeder \cite{sourceforgeOpenDSSCode}, which represents a distribution system
in the U.S. Northeast, with open-source data and models provided by EPRI. Key attributes of the  feeder are summarized in Table \ref{J1_att}.
In addition, J1 has a diverse mix of underground and overhead lines with different impedance characteristics and single-phase loads/transformers ranging from 0.3 kW/5 kVA to 20 kW/75 kVA. With this level of complexity, the phasor-based approach used in the previous section may not yield sufficient insights. Instead, the results from OpenDSS are used to perform the analysis. Since there is no time series data available for the J1 feeder, data from Pecan Street \cite{pecandataport} were applied to the loads and PVs in the J1 feeder to perform the analysis across different time instants. 

\begin{table}[H]
\caption{Attributes of EPRI J1 feeder}
\label{J1_att}
\centering
\renewcommand{\arraystretch}{1.5} 
\resizebox{\columnwidth}{!}{%
\begin{tabular}{|l|c|l|}
\hline
          \textbf{Parameter}                  & \textbf{Value}    & \textbf{Comments}     \\\hline
Primary Voltage   & 12.47 kV   & Substation transformer at 69 kV         \\\hline
        Secondary Voltage                   & 240 V  &       \\\hline
       Total Customers    & 1384    & 363/375/643/3 (Phase A/B/C/3-ph)         \\\hline
       Total Nodes & 4245 & 2037 Primary/2208 Secondary \\\hline
      
   Total Load   & 10.95 MW & Includes 5 MW aggregated load  \\\hline
    Transformers & 819 & 218/225/372/4 (Phase A/B/C/3-ph) \\\hline
   SVRs & 9 & 3/3/2/1 (Phase A/B/C/Substation) \\\hline
   Capacitors & 3900 kVAR & 5 Capacitors \\\hline
   Existing PV & 1813.6 kW & 1.71 MW commercial, 103.6 kW residential \\\hline 
\end{tabular}
}
\end{table}

\subsection{Effect of Reactive Power (Q) Intervention from PV Nodes}
\begin{figure}[htbp]
\centering
  \includegraphics[width=0.485\textwidth]{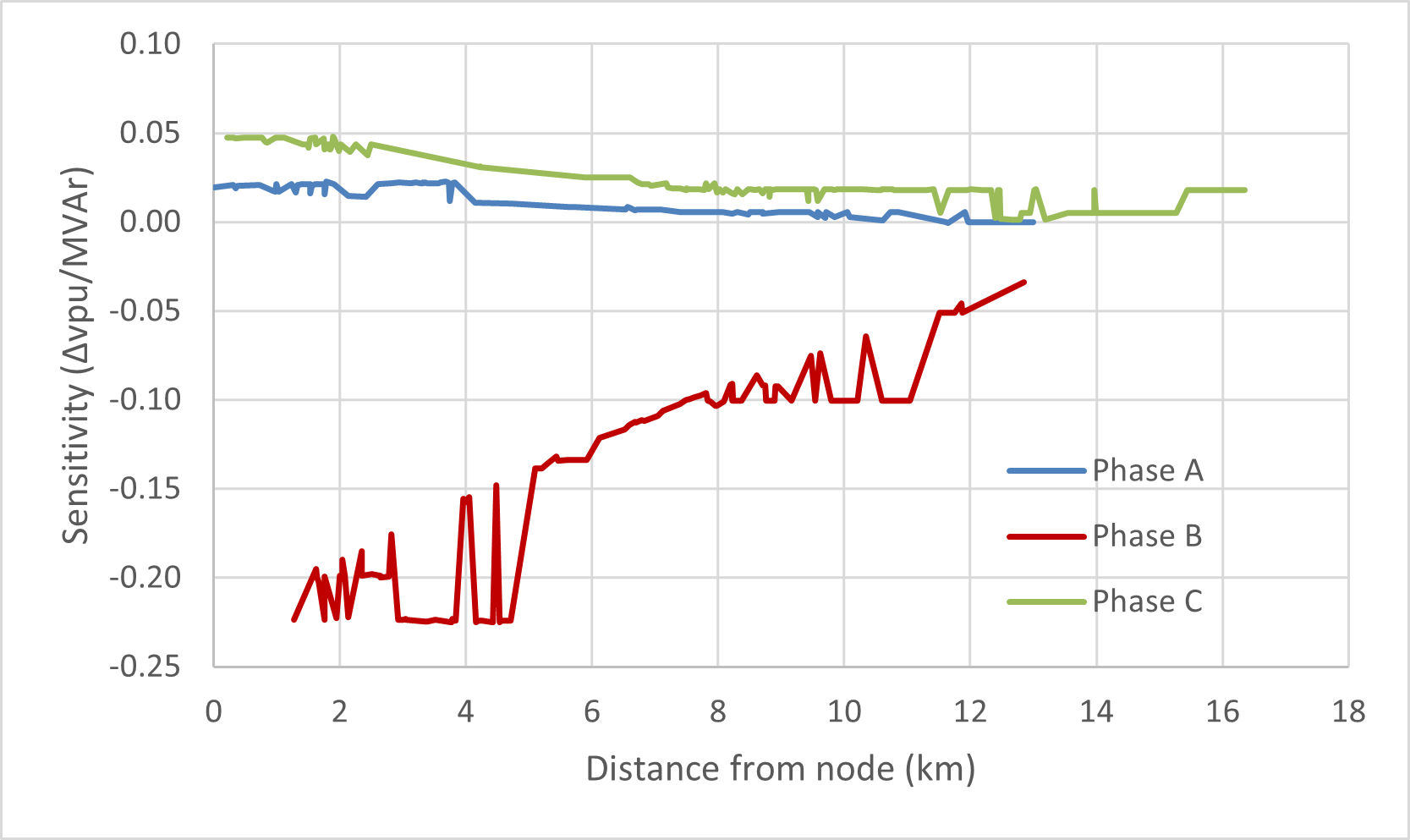}
  \caption{Sensitivity of a single SVR node to all PVs}
  \label{fig: xphase1}
\end{figure}

To
increase 
the J1 feeder HC without incurring voltage violations,
an extreme use case (11 AM, August) was selected. 
For this use case,
a voltage-reactive power sensitivity matrix (VQ-SM) was constructed, with its elements given by \eqref{s_comp}, where $\Delta Q_j$ is the reactive power perturbation applied to PV inverter $j$, 
and $v_i^{0}$ and $v_{i,j}^Q$ are the voltages at node $i$ prior to the perturbation and after the perturbation, respectively.

\begin{equation}\label{s_comp}
    sm_{i,j} = \frac{v_{i,j}^Q - v_i^{0}}{\Delta Q_j}
\end{equation}
Since the VQ-SM is large (rows corresponding to number of PVs and columns corresponding to number of nodes in the system)
we can look at one column which depicts the effect on a single system node from all the PVs where $\Delta Q$ can be applied. It was observed that the secondary node of a single-phase SVR in phase B, B19007reg.2, exhibited highest voltage on a consistent basis under moderate-to-high PV penetration. 
This SVR is located downstream from the feeder-head and has a step-up ratio to restore the voltage levels under high load conditions. 
At 11 AM, when PV generation is high with moderate load conditions, the voltage on the primary side of the SVR goes up (but does not have violations). 
However, due to the step-up ratio, the secondary node and other nodes close to the SVR (including some customer sites) experience voltage violations.  
The method to address these violations is to identify PV nodes that have the highest impact on this SVR secondary node and activate them for Q intervention. Fig. \ref{fig: xphase1} depicts the sensitivity of all the PV nodes in the system to the selected node (B19007reg.2). 

The following observations are made from Fig. \ref{fig: xphase1}:
\begin{itemize}
    \item Sensitivity values for phase B absorption are negative as the node under consideration is a phase B node. Both phase A and phase C sensitivities are positive - a result consistent with the explanation in Section \ref{Simple_Ch}.
    \item The sensitivity magnitudes generally decrease as the distance from the node increases. However, there are clear exceptions indicated by the choppiness of the plots.
    These can be attributed to the fact that the physical distances do not always equate to the electrical distances (impedances). Additional factors such as the presence of CBs and SVRs, and location relative to the substation (upstream or downstream) also contribute to these incongruous results.
    \item  If only the Q interventions of phase B are considered to address the voltage violations, it may not be sufficient. This is because many of the nearby PVs in phases A and C may be activated to address voltage violations in the nodes located on those phases, and they could trigger decremental effects on the selected SVR node. Hence, one must 
    consider all phase effects together to achieve the desired results.   
\end{itemize}

The inferences drawn from these observations are used to implement an empirical prioritized voltage control strategy that is explained in the next sub-section.

\subsection{Prioritized Q Intervention for Effective Voltage Control}
\label{manual}
An empirical Algorithm \ref{Alg_1} is now devised to add PVs into a distribution system for increased HC while accounting for the cross-phase effects explained previously.
The main purpose of this algorithm is to provide a relatable illustration of the cross-phase effects and the means to overcome them. 

\begin{algorithm}[H]
\caption{PV addition/Voltage control algorithm} \label{Alg_1}
\begin{algorithmic}[1]
\STATE Load network details into a distribution system solver
\STATE Add $n$ PV system, where $n=1$
\STATE Apply appropriate PV and load profiles
\STATE Select evaluation condition (time instance)
\STATE Run power flow with unity PF (UPF) and identify locations and values of voltage violations by phase

\STATE Identify PVs with highest sensitivity to the UPF violation nodes and incrementally add $\Delta Q$ on them till zero violations are achieved in power flows
\STATE Repeat Steps 3 to 6 with $n=n+1$ PVs
\end{algorithmic}
\end{algorithm}

\begin{figure}[htbp]
\centering
  \includegraphics[width=0.48\textwidth]{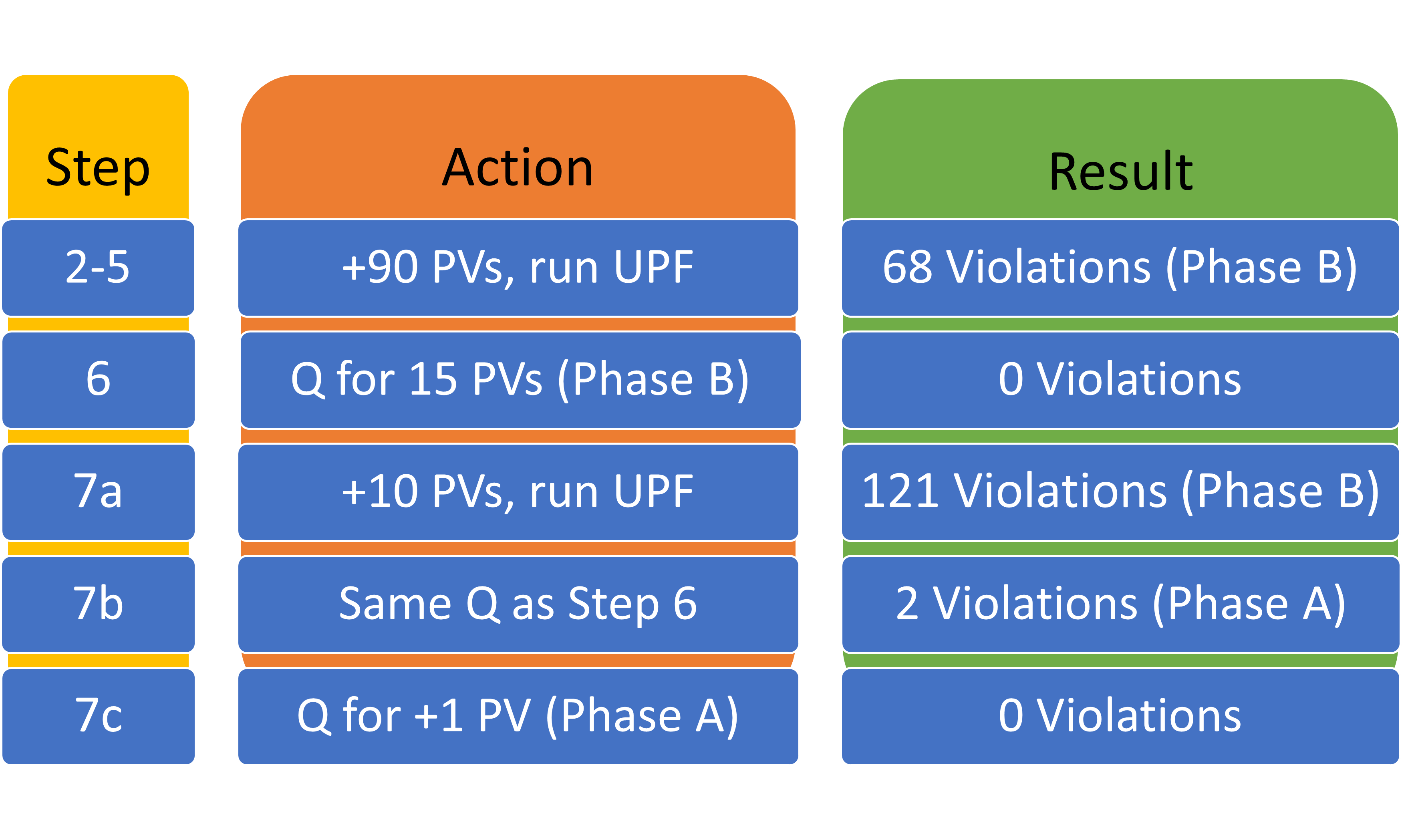}
  \vspace{-5mm}
  \caption{Cross-phase sensitivity effects using Algorithm \ref{Alg_1}}
  \label{fig: xphase2}
\end{figure}

The results of running Algorithm \ref{Alg_1} on EPRI J1 feeder are depicted in Fig. \ref{fig: xphase2}. These results clearly identify the decremental effects of uncoordinated Q intervention. 
As shown in the first two rows, for 90 PV additions, all violations are in phase B (68 of them), and they were addressed (Step 6) by activating Q intervention in a prioritized list of 15 PVs in phase B, guided by the VQ-SM.
When 10 more PVs were added (Step 7a), there were more phase B violations (121 of them), but no phase A or phase C violations under UPF. 
However, when the Q intervention from Step 6 was activated (Step 7b), there were two new violations in phase A. The dominant contributor to the new violations in phase A was the decremental effect from Q intervention on the 15 phase B PVs. 
This was further confirmed by reducing the Q intervention in phase B, which removed the phase A violation, but reintroduced the phase B violations. The final solution, as shown in the last row (Step 7c), was to add Q intervention at one phase A PV. 

\section{Impact on Voltage Control Strategies}
\label{VCstrategies}
The simple example in Section \ref{manual} clearly showed the decremental effect of cross-phase sensitivity. Looking at the larger picture, it also points to the challenges in achieving effective coordinated voltage control with Q intervention in real-world, large, complex distribution systems. 
For example, with any partition that is phase-based or distance-based, there will be a significant loss of accuracy that can lead to sub-optimal results from a voltage control perspective.

\begin{figure}[htbp]
\centering
  \includegraphics[width=0.48\textwidth]{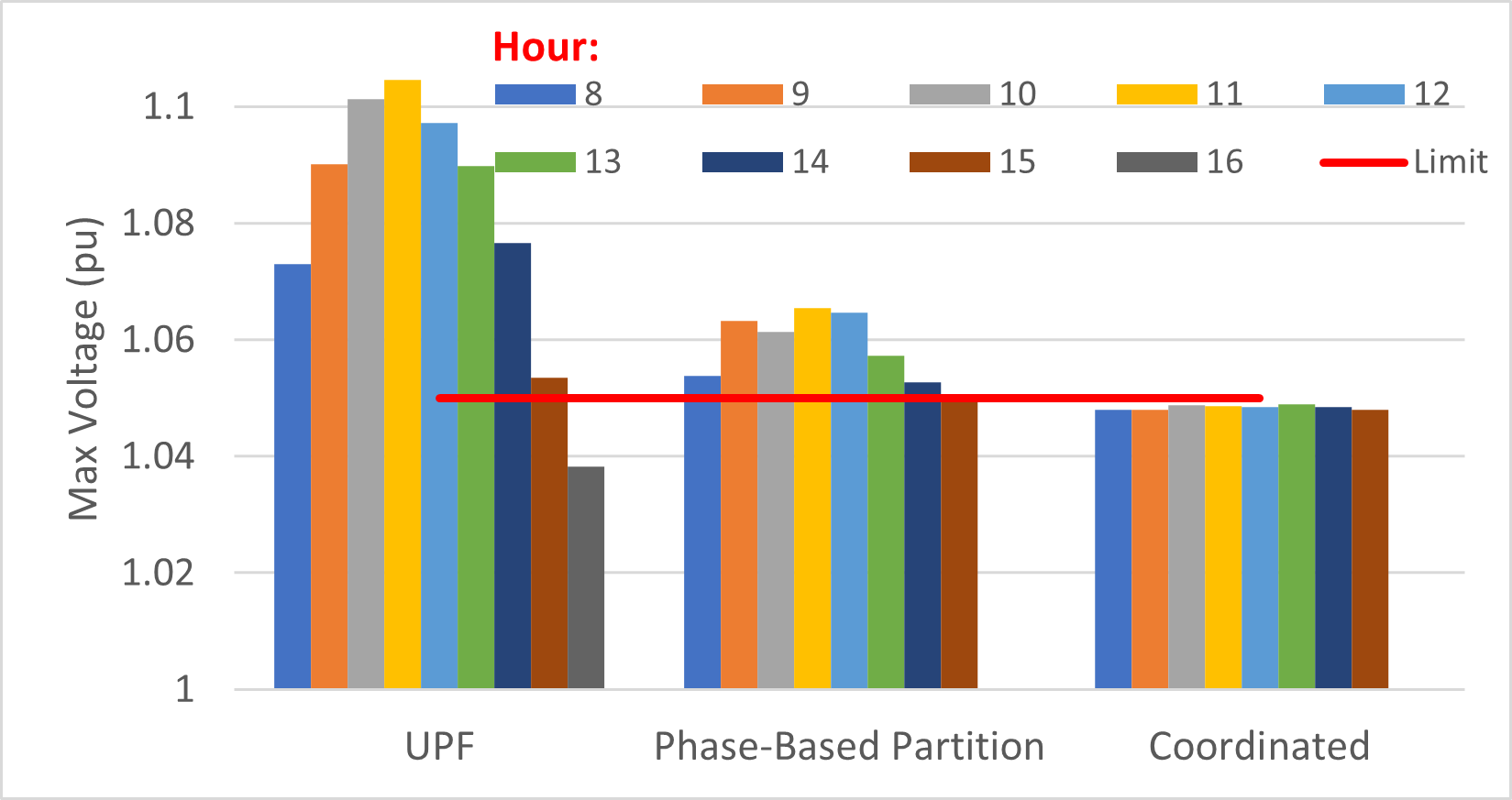}
  \caption{Cross-phase sensitivity effects with high PV penetration}
  \label{fig: Res3ph}
\end{figure}

In order to demonstrate this impact, a large number (460) of residential PVs were added in a random fashion to the EPRI J1 feeder, taking the total PV capacity to 6.4 MW. With the inverters set at UPF (i.e., without any Q intervention), there were many voltage violations for different hours of the day with the maximum voltage exceeding 1.1 p.u. during some hours as shown in the left part of Fig. \ref{fig: Res3ph}. 

Next, the network is partitioned into three zones, one for each phase. For each phase, a VQ-SM is created using only the PVs and nodes in that phase, and three optimization algorithms are run independently of each other for the three phases. 
The goal of each of the optimization algorithms was to find the minimum Q intervention in a given phase to minimize the voltage violations of that phase. 
The results from the three algorithms are combined and applied to the distribution system solver (OpenDSS) 
to check for voltage violations.
The results in the middle part of Fig. \ref{fig: Res3ph} clearly depict the relative ineffectiveness of the phase-based partitioning. By omitting the cross-phase sensitivity from the optimization formulation, the optimizer is misled into providing Q intervention solutions that may work for one phase, but not for the other phases.

Finally, an optimization-based iterative voltage control algorithm was developed to mitigate the violations, while also accounting for the cross-phase sensitivity effects. This algorithm relied on availability of system-wide voltage information and iterative refinement of the VQ-SM for its effectiveness. More details about this algorithm can be found in \cite{dalal2023improving}.
As shown in the right part of Fig. \ref{fig: Res3ph}, this algorithm is successful in removing voltage violations at all hours, which is impressive considering the extent of violations in the UPF case in terms of numbers (3000+) and voltage magnitudes (1.1045 p.u.). 

\section{Conclusion}
This paper presented a robust argument for considering the cross-phase effects of Q intervention in distribution systems to achieve better voltage control and higher levels of HC. The underlying reasons for the cross-phase sensitivity were identified first.
Then, a simple network model was created and phasor diagrams were employed to analytically depict how the cross-phase effects manifest in the system. 

Next, successive illustrations 
on a complex real-world distribution feeder underlined how the cross-phase effects negatively impact voltage control algorithms if they are not fully considered within the control strategy. Finally, 
a demonstration of
how the incorporation of these effects in the control algorithms lead to significantly better results was provided.

{\footnotesize
\bibliographystyle{IEEEtran}
\bibliography{IEEEabrv, references}
}

\end{document}